\documentstyle[12pt]{article}
\begin{document}

\begin{center}
{\bf Classical electromagnetism of $r^2$ spin-orbit interaction and 
nonexistence of magnetic monopoles}
\end{center}

\bigskip

\begin{center}
H.C. Rosu {\scriptsize and} F. Aceves de la Cruz


{\scriptsize Instituto de F\'{\i}sica, Univ. de Guanajuato, Apdo Postal E-143, 
Le\'on, Gto, 
Mexico}
\end{center}

\bigskip
\bigskip

{\bf Abstract} - We first obtain the electric and magnetic fields 
corresponding to a `spin'-orbit classical interaction of a $r^2$ potential. 
Assuming
that Maxwell equations hold for these fields, we infer the conditions on the
`spin' vector forbidding magnetic monopoles.

\bigskip 
\bigskip 
\bigskip 
\bigskip 
\bigskip

Let us consider Maxwell's equations (symbols are standard ones)
\begin{eqnarray*} 
&&\nabla\times\vec{H} - \frac{\partial\vec{D}}{\partial{t}} = 
\frac{4\pi}{c}\vec{j}\\ 
&&\nabla \cdot \vec{B} = 0,\\
&&\nabla\times\vec{E} + \frac{1}{c}\frac{\partial\vec{B}}{\partial{t}}=0,\\
&&\nabla \cdot \vec{D} = 4\pi\rho ~.
\end{eqnarray*}

The Lorentz force reads
\begin{eqnarray*}
\vec{F} &=& q(\vec{E} + \frac{1}{c}\vec{v} \times \vec{B})\\
&=& -\nabla{U} + \frac{d~}{dt}\frac{\partial{U}}{\partial\vec{v}}~,
\end{eqnarray*}
where $U = q\Phi - (q/c)\vec{A} \cdot \vec{v}$ (see Ref. 1). 
We are interested in the following case 
\[
U = V(r) + \vec{\sigma} \cdot \vec{L}~,
\]
which is similar to a spin-orbit interaction for $V(r)\propto r^2$, though we
shall consider the `spin' vector as an arbitrary classical vector. 
One can write
the spin-orbit term as $-m(\vec{r} \times 
\vec{\sigma}) \cdot \vec{v}$. Comparing with the general case, we get
\[
\vec{A} = \frac{mc}{q}\vec{r} \times \vec{\sigma}.
\]


On the other hand, it is known that 
\begin{eqnarray*}
\vec{E} = -\nabla\Phi - \frac{1}{c}\frac{\partial\vec{A}}{\partial{t}},\\
\vec{B} = \nabla \times \vec{A}
\end{eqnarray*}
and taking $\Phi = V(r)/q$ leads to
\begin{eqnarray*}
&&\vec{E} = -\frac{1}{q}\left(\nabla{V(r)} + 
m\frac{\partial~}{\partial{t}}(\vec{r} \times \vec{\sigma})\right)\\
&&\vec{B} = \frac{mc}{q}[(\nabla \cdot \vec{\sigma})\vec{r} - 3\vec{\sigma}]~.
\end{eqnarray*}

Plugging these results into Maxwell's equations gives
\begin{eqnarray*}
\nabla \cdot \vec{B} = 0 &\rightarrow& \vec{r} \cdot \nabla(\nabla \cdot 
\vec{\sigma}) = 0\\
\nabla \cdot \vec{D} = 4\pi\rho &\rightarrow& \rho = 
-\frac{\epsilon}{4\pi{q}}\left(\nabla^2V + 
m\frac{\partial~}{\partial{t}}(\vec{r} \cdot 
\nabla\times\vec{\sigma})\right)\\
\nabla\times\vec{H} - \frac{\partial\vec{D}}{\partial{t}} =
\frac{4\pi}{c}\vec{j} &\rightarrow& \vec{j} = 
\frac{m\epsilon}{4\pi{q}}\frac{\partial^2~}{\partial{t^2}}(\vec{r} 
\times \vec{\sigma}) - \frac{3mc^2}{4\pi\mu{q}}\nabla\times\vec{\sigma}~.
\end{eqnarray*}
The first of these results implies that nonexistence of magnetic monopoles 
requires $\vec{\sigma}$ be only a linear or constant function of the 
coordinates 
with arbitrary explicit time dependence.

\bigskip
\bigskip

Ref. 1 - H.C. Rosu, physics/0005019

\end{document}